\documentstyle[psfig]{mn}

\newif\ifAMStwofonts

\ifoldfss
  \ifCUPmtlplainloaded \else
    \NewTextAlphabet{textbfit} {cmbxti10} {}
    \NewTextAlphabet{textbfss} {cmssbx10} {}
    \NewMathAlphabet{mathbfit} {cmbxti10} {} 
    \NewMathAlphabet{mathbfss} {cmssbx10} {} 
  \fi
  \ifAMStwofonts
    \ifCUPmtlplainloaded \else
      \NewSymbolFont{upmath} {eurm10}
      \NewSymbolFont{AMSa} {msam10}
      \NewMathSymbol{\upi}     {0}{upmath}{19}
      \NewMathSymbol{\umu}     {0}{upmath}{16}
      \NewMathSymbol{\upartial}{0}{upmath}{40}
      \NewMathSymbol{\leqslant}{3}{AMSa}{36}
      \NewMathSymbol{\geqslant}{3}{AMSa}{3E}

    \fi
  \fi
\fi 

\ifnfssone
  \newmathalphabet{\mathit}
  \addtoversion{normal}{\mathit}{cmr}{m}{it}
  \addtoversion{bold}{\mathit}{cmr}{bx}{it}
  \newmathalphabet{\mathbfit} 
  \addtoversion{normal}{\mathbfit}{cmr}{bx}{it}
  \addtoversion{bold}{\mathbfit}{cmr}{bx}{it}
  \newmathalphabet{\mathbfss} 
  \addtoversion{normal}{\mathbfss}{cmss}{bx}{n}
  \addtoversion{bold}{\mathbfss}{cmss}{bx}{n}
  \ifAMStwofonts
    \ifCUPmtlplainloaded \else
      \UseAMStwoboldmath
      \makeatletter
      \new@mathgroup\upmath@group
      \define@mathgroup\mv@normal\upmath@group{eur}{m}{n}
      \define@mathgroup\mv@bold\upmath@group{eur}{b}{n}
      \edef\UPM{\hexnumber\upmath@group}
      \new@mathgroup\amsa@group
      \define@mathgroup\mv@normal\amsa@group{msa}{m}{n}
      \define@mathgroup\mv@bold\amsa@group{msa}{m}{n}
      \edef\AMSa{\hexnumber\amsa@group}
      \makeatother
      \mathchardef\upi="0\UPM19
      \mathchardef\umu="0\UPM16
      \mathchardef\upartial="0\UPM40
      \mathchardef\leqslant="3\AMSa36
      \mathchardef\geqslant="3\AMSa3E
    \fi
  \fi
\fi 

\ifnfsstwo
  \DeclareMathAlphabet{\mathbfit}{OT1}{cmr}{bx}{it}
  \SetMathAlphabet\mathbfit{bold}{OT1}{cmr}{bx}{it}
  \DeclareMathAlphabet{\mathbfss}{OT1}{cmss}{bx}{n}
  \SetMathAlphabet\mathbfss{bold}{OT1}{cmss}{bx}{n}
  \ifAMStwofonts
    \ifCUPmtlplainloaded \else
      \DeclareSymbolFont{UPM}{U}{eur}{m}{n}
      \SetSymbolFont{UPM}{bold}{U}{eur}{b}{n}
      \DeclareSymbolFont{AMSa}{U}{msa}{m}{n}
      \DeclareMathSymbol{\upi}{0}{UPM}{"19}
      \DeclareMathSymbol{\umu}{0}{UPM}{"16}
      \DeclareMathSymbol{\upartial}{0}{UPM}{"40}
      \DeclareMathSymbol{\leqslant}{3}{AMSa}{"36}
      \DeclareMathSymbol{\geqslant}{3}{AMSa}{"3E}
    \fi
  \fi
\fi 

\ifCUPmtlplainloaded \else
  \ifAMStwofonts \else 
    \def\upi{\pi}
    \def\umu{\mu}
    \def\upartial{\partial}
  \fi
\fi

\title{Flaring and warping of the Milky Way disk: not only in the gas....}
\author[C. Alard]
       {C. Alard \\
        Institut d'Astrophysique de Paris, 98bis Boulevard Arago, F-75014.}
\date{}

\pagerange{\pageref{firstpage}--\pageref{lastpage}}
\pubyear{2000}

\begin{document}

\maketitle

\label{firstpage}

\begin{abstract}
 This paper presents an investigation of the outer disk structure
 by using data from a recent release of the 2 micron sky survey (2MASS).
 This 2MASS data show unambiguously that the stellar disk thickens
 with increasing distance from the Sun. In one of the field (longitude l$=$240) 
 there is also strong evidence of an assymetry associated with the Galactic warp. 
 This flaring and warping of the stellar disk is very
 similar to the features observed in the HI disk. The thickening of the stellar disk
 explains the drop in density observed near the Galactic plane: stars located at 
 lower Galactic latitudes are re-distributed to higher latitudes. It is no longer
  necessary to introduce a disk cut-off to explain the drop in density. It is
  also not clear how this flaring disk is distinct from the thick disk in the outer 
 disk region.  At least, for lines of sight in the direction of the outer
 disk, the thickening of the disk is sufficient to account
 for the excess in star counts attributed by some models to the thick disk. 
\end{abstract}

\section{Introduction.}
The structure of the stellar disk of our Galaxy is still not very well known.
 The stellar population of the disk is usually decomposed in a thin disk and thick
 disk component. The thin disk and thick disk populations are separated
 on the basis that they may have  different metallicity and velocity profiles.  
 The typical ratio of the thick disk to thin disk components in the solar 
 neighborhood is about a few percent (Gilmore \& Reid 1983).
 . 
 Both disk are supposed to follow a double exponential profile, associated with
 a radial scale length and a vertical scale height.
 It is unclear at the moment if the
 scale length of these 2 disks are the same. There is still some uncertainty 
 even for the scale length of the thin disk itself, although the more recent
 measurement seems to indicate that it is between 2 and 3 Kpc. For the the thick
 disk the uncertainty is much larger, typically the scale length is in the range
 1.5 to 4.5 Kpc. The scale heights of the thin disk is about 0.3 Kpc ,
 and is around 1 Kpc for the thick disk.
 The structure of the gas component (whether HI or CO) is better understood, and
 suggests some questions concerning the stellar disk. Since a warp is well
 identified in the gas ( W. B. Burton and P. te Linkel Hekkert 1986), one may wonder 
 if it does exists also in the stellar components, with the same amplitude and phase.
 Another problem is the extension of the gas component which is visible up to a distance
 of 25 Kpc from the Galactic center. Some observations suggests that the stellar disk
 ends long before, at about 14 Kpc from the center. This fact may indicate that
 stellar formation in the disk did not occur beyond 14 Kpc, or that the geometry
 of the stellar disk is not well understood. One additional source of complication
 is the possibility that the disk in addition to its warp mode presents an elliptical
 mode of deformation. At the moment this issue is very speculative, however this
 effect may exist and should be investigated.
\section{The data}
\subsection{Presentation}
 The data set consist of 3 strips perpendicular to the Galactic plane around the longitudes,
 l$=$66, l$=$180, l$=$240. All the data have been extracted form the 2MASS point source
 catalogue accessible on the Web (http://www.ipac.caltech.edu/2MASS/). 
The detailed characteristics of the fields are given 
 in Table 1.
\begin{table} 
   \caption{The 3 fields}
   \label{}
   \begin{tabular}{llll}
      \hline\noalign{\smallskip}
   Field & field center & size of field & Number of stars\\ 
   \hline\noalign{\smallskip}
     L66 & (l,b)$=$(66,0) & ($\Delta$l,$\Delta$b)$=$(2,100)  &  \ \ 1.4 $10^6$ \\ 
     L180 & (l,b)$=$(180,0) & ($\Delta$l,$\Delta$b)$=$(10,100) &  \ \ 3.8 $10^6$ \\  
     L240 & (l,b)$=$(240,0) &  ($\Delta$l,$\Delta$b)$=$(10,100) &  \ \ 4.7 $10^6$  \\
   \hline\noalign{\smallskip}
      \noalign{\smallskip} 
  \end{tabular}
\end{table}

\subsection{Canceling the effect of extinction.}
 By combining 2 photometric bands it is easy to derive a magnitude that is independent of
 reddening. For instance using J and K, we can derive $M_E$:
$$
 M_E = K - \frac{A_K}{A_J-A_K} \ (J-K)
$$
 Following  Rieke and Lebofsky (1985) we take: 
 $$
 \frac{A_K}{A_V} = 0.112 \ \ \ {\rm and}  \ \ \  \frac{A_J}{A_V} = 0.282 \ \ \ \ \ \ \ \  
 $$ 
Leading finally to:
\begin{equation}
 M_E = K - 0.659 \ (J-K)
\end{equation}
This corrected magnitude will be very insensitive to reddening, furthermore, it is important
 to notice that the reddening in the 3 fields is low (see Fig. 1). The maximum reddening
 is about $A_V \simeq 1.5$, which is only 0.17 in the K band. Consequently, after
 adding the correcting term: $-0.659 \ (J-K)$, the residual will not exceed a few percent.
\begin{figure} 
\centerline{\psfig{angle=0,figure=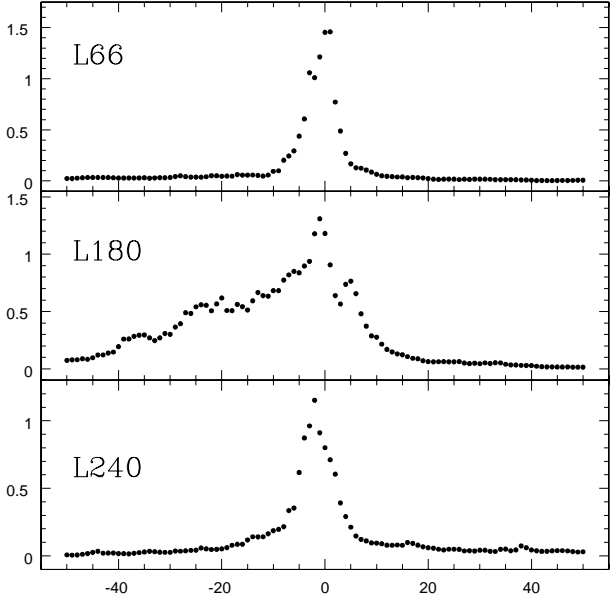,width=9cm}}
\caption{Mean extinction in the V band in the 3 fields. These plot were constructed
 from the map of Schlegel {\it et al.} (1998).
}
\end{figure}
\subsection{Star counts as a function of latitude}
Before making the star counts, it is essential to correct for empty areas in the catalogue
 (due most of the time to the obliteration of fainter stars by a very bright star). To locate
 these bad areas a 2 dimensional histogram in the l and b coordinates was computed. The size
 of the bin was chosen in order to have a mean of 50 stars in the bin. It was required that
 the bin contains a minimum of 5 stars to be considered valid. This way the histogram of the
 valid areas was constructed as a function of latitude. The step in latitude is 0.5 degree.
 The counts were finally normalized to a surface of 1 square degree. To avoid any problem
 related to the completeness of the sample, only stars brighter than $M_E=12.5$ have been 
 included. The resulting histograms are presented in Fig. 2. 
\begin{figure} 
\centerline{\psfig{angle=0,figure=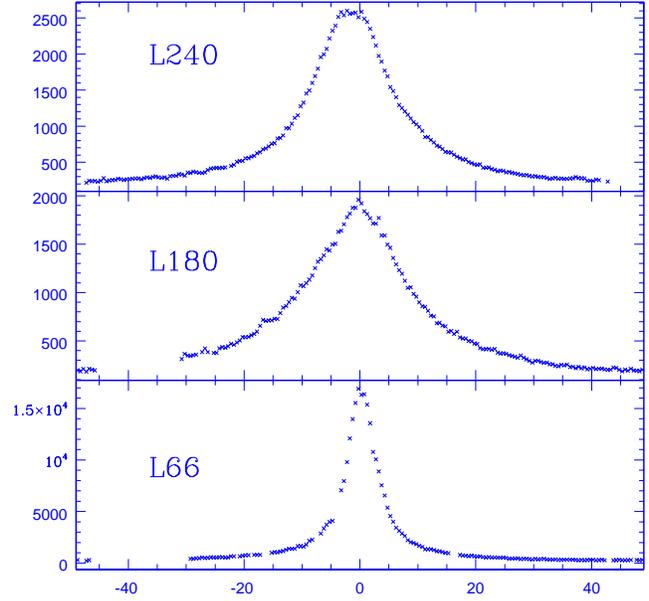,width=9cm}}
\caption{
 Star counts for stars brighter than $M_E=12.5$
 as a function of latitude for the 3 fields. The counts
 are normalized in number of stars per square degree. 
  }
\end{figure}
\begin{figure} 
\centerline{\psfig{angle=0,figure=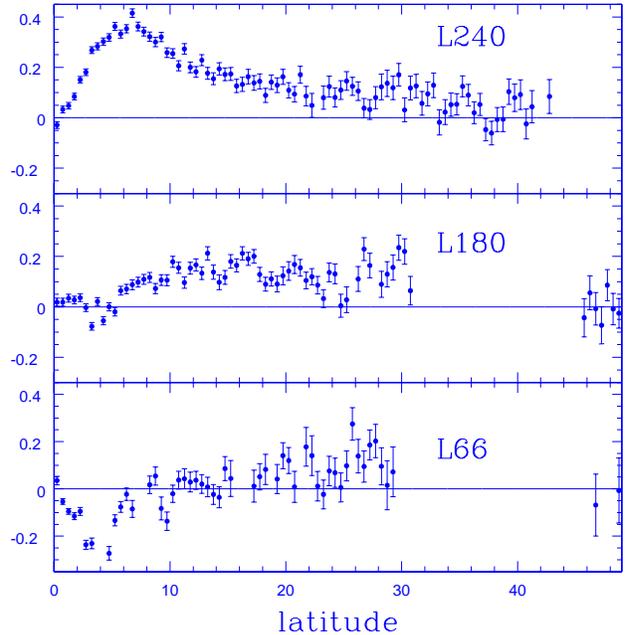,width=9cm}}
\caption{
 Difference (percentage) between counts at positive and negative latitude as a function of
 latitude.
  }
\end{figure}
\subsection{Assymetry between positive and negative longitude.}
It is interesting to investigate the symmetry of the distribution
 of the counts as a function of latitude. A significant assymetry would reveal
 the presence of the Galactic warp. In Fig.3 the difference in star
 counts between positive and negative longitude is plotted. It appears
 immediately that the field L240 presents large deviations from symmetry
 near b$=$6. It is likely that this assymetry is associated
 with the presence of a significant Galactic warp in this field. Note that possible
 evidences for an assymetry in the star counts has already been presented by
 Carney \& Seitzer (1993).
 For the 2 other fields the pattern of assymetry is less clear.
 For l=180 we observe some systematic deviation near the 2 $\sigma$ level, in
 the range b$=$10 to  b$=$30. For the field at l$=$66 there are a few
 deviating points near b$=$4. 
\section{Analyzing the data.}
\subsection{Deviations from the exponential profile.}

To estimate numerically the star counts, we need
 to integrate the luminosity function of the stellar population and the density
 function of the disk along the line of sight. The luminosity function is assumed
 to be the same everywhere and is identical to the luminosity 
 function in the solar neighborhood tabulated by Wainscoat {\it et al.} (1992).
 To represent the density profile of the disk, it is natural 
 to look first at the simplest model, which is a single double exponential:
$$
 \rho(R,Z) = \exp{\left(-\frac{R-R0}{R_h}-\frac{|Z|}{z_h} \right)} \ \ \ with \ \ R=\sqrt{(x^2+y^2)}
$$
 It is important to note than despite its simplicity, the exponential
 profile seems to show en excellent consistency with the Z-profile
 of edge-on galaxies within a few scale length form the center (de Grijs {\it et al.} 1996).
 It is interesting to check the consistency of the data with such simple model, 
 and in particular to study the shape of the deviations from this model which may indicate
  that other components are present. It is  possible to study this model in a the full
 range of latitude covered by our data ($-50<b<50$, 169 line of sight).
By probing different line of sight, with increasing latitudes,
 we probe areas of the disk more and more distant form the Galactic plane.
 Thus basically, by doing this, we explore the $Z$ profile of the 
 Galactic disk. If the disk is really a 
 double exponential, this $Z$ profile should be a constant exponential, 
 whatever the distance $R$ from the center. Consequently along all the line
 of sight the density should be consistent with an exponential in $Z$
 with constant scale height, $H_Z$. In case the $Z$ profile is not
 exponential we should observe a variation of the scale height as a function
 of the latitude. To first order in $Z$ we can develop the variation to
 the $Z$ profile for each latitude $b$ is equivalent to a variation
 of the scale height $H_Z$:
$$
 \rho = \exp{\left(-\frac{R}{R_h}-\frac{|Z|}{Z_h} -f(b) \ |Z|\right)} = 
\exp{\left(-\frac{R}{R_h}-\frac{|Z|}{H_Z(b)}\right)} 
$$
 it is easy to reconstruct the function $H_Z(b)$ for the 169 line of sight. Along
 each line of sight the data have been decomposed in 20 bins of magnitudes,
 allowing direct maximum likelihood determination of $H_Z(b)$
 by maximum likelihood. It is interesting to note that the estimation of 
 the factor $H_Z(b)$ along each line site can also compensate for an imperfect
 estimation of the scale length $R_h$. This is obvious for the line of sight
 near $l=180$, since the distance $d$ is proportional to Z in this direction.
 For the line of sight near $l=240$ this is still a good approximation. 
 An even for $l=66$, one can check easily numerically that the factor $H_Z(b)$ 
 can compensate with good accuracy for small discrepancies in the estimation of $R_h$.
 The function $H_Z(b)$ has been estimated for the 3 fields, allowing a full reconstruction
 of the density profile of the disk. In order to make an easy comparison
 with an exponential profile, the Z profile of this density has been computed
 for different values of the distance to the center $d$. The result for the 3 fields
 is presented in Fig. 4,5,6. The most obvious difference between the 3 fields is that
 the vertical profile does not deviate much from the straight line (pure exponential)
 all along the line of sight in the field at $l=66$. The situation is very different 
 in the 2 other fields, where the profile becomes broader and broader with distance
 with an increasingly large deviations from an exponential. One could always claim that
  this is a bias due to the fact that the Z profile is constant
 but not exponential. However, we see immediately that this is not possible, since in this
 case the same discrepancy should be observed in the field at $l=66$. There are small
 deviations near the top of the profile at $l=66$, which could be due to the fact that we reach
 a distance of about 10 Kpc, for which we already observe some flattening in the other fields.
 It might be also that in this field, source confusion affects the star counts in a small in
 a small range of latitude ($-1.5<b<1.5$). For the 2 other fields, the stellar density is about
 10 times smaller, thus we do not expect any significant effects due to the source confusion. 
\begin{figure} 
\centerline{\psfig{angle=0,figure=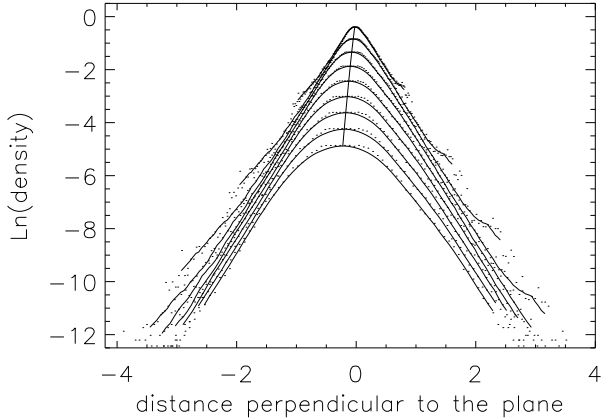,width=9cm}}
\caption{
 The vertical profile of the disk for the field at $l=240$. The profile has been reconstructed
 for distances ranging from 1 to 10 Kpc along the line of sight at $b=0$ (according
 to the limiting magnitude and the luminosity function, the sensitivity
 of the data beyond 10 Kpc is very limited). The corresponding
 distance from the Galactic center are: $R=$ 8.5,9.2,9.8,10.6,11.4,12.2,13.0,13,9,14.7 Kpc.
  The deviation from the exponential profile become very large
 beyond 13 Kpc. The solid lines are computed by smoothing of the data points. 
 A line passing by the maxima of the different profiles has also been represented. Note that this
 line is not vertical, which indicates that the distribution is not symmetrical, and most likely
 warped.
  }
\end{figure}

\begin{figure} 
\centerline{\psfig{angle=0,figure=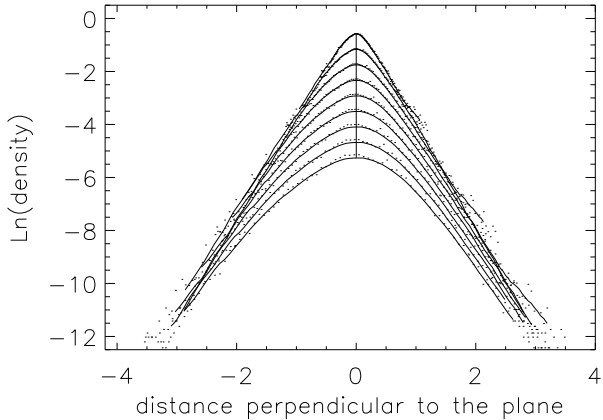,width=9cm}}
\caption{
 The vertical profile of the disk for the field at $l=180$. The profile has been reconstructed
 for distances ranging from 1 to 9 Kpc along the line of sight at $b=0$. The corresponding
 distance from the Galactic center are: $R=$ 9,10,11,12,13,14,15,16,17 Kpc. The flattening of
 the profile with distance is obvious. The deviation from the exponential profile become very large
 beyond 13 Kpc.
 A line passing by the maxima of the different profiles has also been represented. Note that this
 line is very close to vertical, which indicates shows that on the contrary to the field near
 $l=240$ the warp may not be present in this field.
  }
\end{figure}
\begin{figure} 
\centerline{\psfig{angle=0,figure=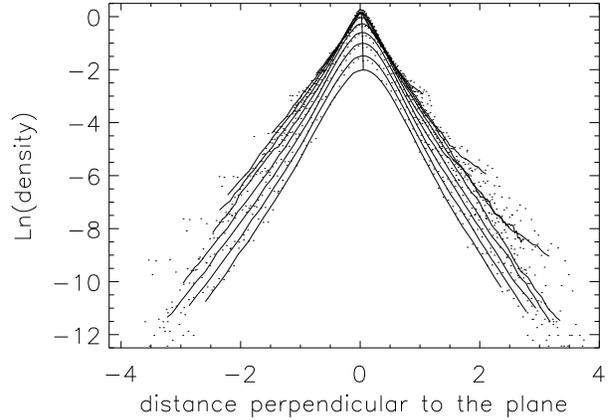,width=9cm}}
\caption{
  The vertical profile of the disk for the field at $l=66$. The profile has been reconstructed
 for distances ranging from 1 to 9 Kpc along the line of sight at $b=0$. The corresponding
 distance from the Galactic center are: $R=$ 7.7,7.4,7.3,7,35,7.5,7.8,8.2,8.7,9.3 Kpc.
 It is important to note that the deviation from the exponential profile are much smaller
 in this field than in the 2 previous fields. It can be explained easily, since in this field
 , even by going 10 Kpc along the line of sight, we do not go beyond 10 Kpc from the Galactic Center. 
  }
\end{figure}
\subsection{Disk models with more parameters.}
It is important to check that this thickening of the disk with distance that we observed is not
 related to some bias, due for instance that we did not included a thick disk. It is easy to check
 this issue by introducing a thick disk component in the model, and to reproduce the fitting
 procedure and the set of vertical profile in this case. Some recent determination of the thick 
 disk parameters indicates that the thick disk accounts for about 0.06 \% of the local density
 (Buser {\it et al.} 1998).
 The scale height of the thick disk is close to 1 Kpc while its radial scale length, is about 
 3 Kpc (Buser {\it et al.} 1998). This thick disk component was included in the model for the field towards 
 the anti-center. 
 The direction of the anti-center for chosen for reason of simplicity. In this field the
 amplitude of the warp is negligeable, and if the disk is elliptical, the effect of ellipticity
 is constant along the line of sight. The resulting set of density profiles is presented 
 in Fig. 7. The addition of the thick disk component does not modify much the shape
 of the profiles. The only effect is only a reduction of the density in the wings of the profiles,
 but it does not affect the flattening. Other model of thick disk with larger density
 or different scale length do not modify the result. This can be easily understood, since due
 to its ``thickness'', the thick disk becomes really important beyond Z $=$ 1.5, and most
 of the flattening effect we observe occur at Z $<$ 1.5.
 Another possibility that has been used to reproduce the the star counts towards the
 anti-center is to used a cut-off of the disk to some distance of the center. For instance
 Robin {\it et al.} (1992) found that a cut-off of the disk at 14 Kpc was necessary in
 order to account for the stellar density in a field close to anti-center direction, 
 near b$=$2.5. The density reconstructed with this cut-off in our model of the disk
 is presented if Fig. 8. This cut-off does not help to reduce the variation of profile
 with distance. This fact was easily predictable, since if we try estimate the cut-off
 for different line of sight, the value of the cut-off is not constant (see Fig. 9 and 10). It shows
 that a cut-off of the disk is not a good solution. This we can conclude that the effect
 of flattening of the profile with distance, is not a bias due to the simplicity of our 
 approach, but a real effect.
\begin{figure} 
\centerline{\psfig{angle=0,figure=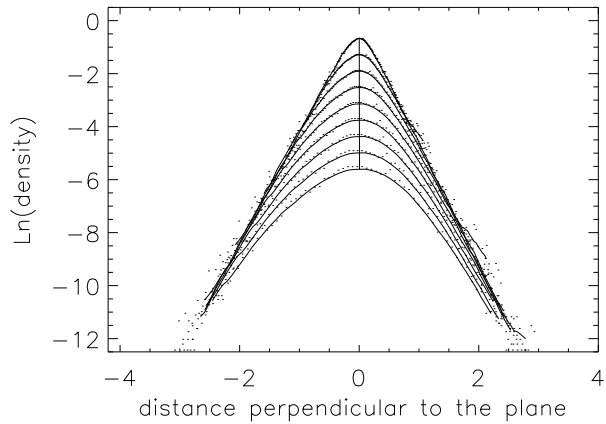,width=9cm}}
\caption{
 The vertical profile of the disk reconstructed using a model including
 a thick disk component (the field is at $l=180$). The profile has been reconstructed
 for distances ranging from 1 to 9 Kpc along the line of sight at $b=0$. The corresponding
 distance from the Galactic center are: $R=$ 9,10,11,12,13,14,15,16,17 Kpc. Note that
 the addition of the thick disk component has some influence in the range Z $=$1.5,2.5. However
 it does not change the shape of the profiles for Z $<$ 1.5, and the basic features
 associated with the flattening of the profile with distance are unchanged.}
\end{figure}
\begin{figure} 
\centerline{\psfig{angle=0,figure=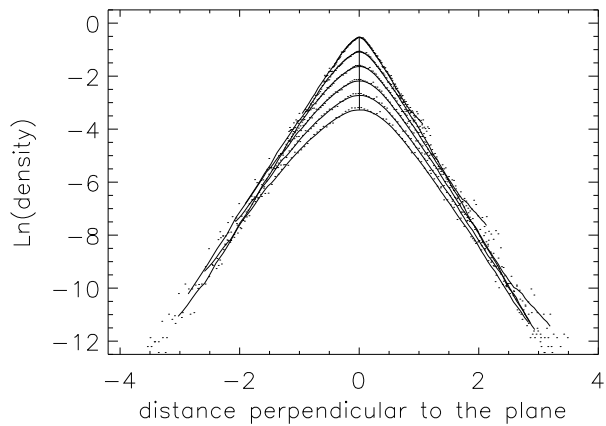,width=9cm}}
\caption{
 The vertical profile of the disk reconstructed in the case of a model
 which includes a cut off of the disk density a $R=14$ Kpc.
 (the field is at $l=180$). The profile has been reconstructed
 for distances ranging from 1 to 9 Kpc along the line of sight at $b=0$. The corresponding
 distance from the Galactic center are: $R=$ 9,10,11,12,13,14 Kpc. Note that
 the effect of the cut-off does not help to reduce the variations of the vertical profile
 with the distance from the center.}
\end{figure}
\begin{figure} 
\centerline{\psfig{angle=0,figure=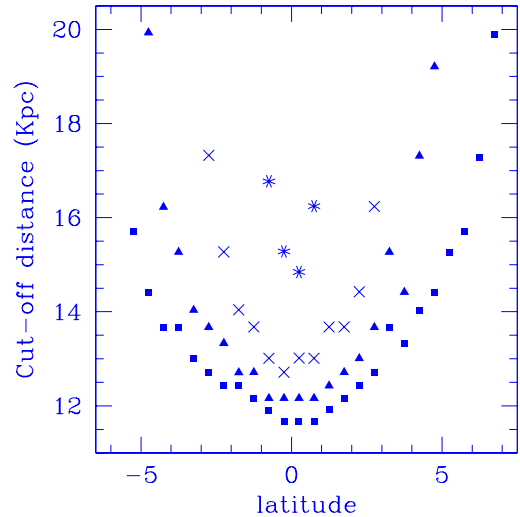,width=7.5cm}}
\caption{
 The cut-off near the Galactic along different line in the field at l$=$180. 
 The cut-off was estimated using
 a maximum likelihood method for different values of the disk scale length:
 3.5 Kpc (squares), 3.0 Kpc (triangles), 2.5 Kpc (crosses), and 2.0 Kpc (stars).
 Note that whatever the scale length the cut-off is extremely variable. Thus one can
 conclude that a disk with cut-off is not appropriate to represent the stellar density
 along the line of sight.}
\end{figure}
\begin{figure} 
\centerline{\psfig{angle=0,figure=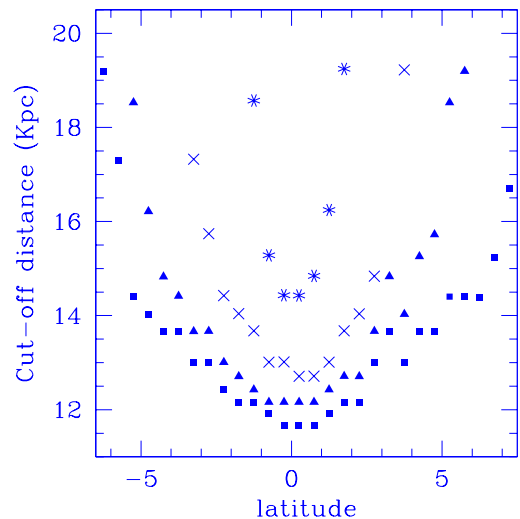,width=7.5cm}}
\caption{
 Same as Fig. 9 except a thick disk component (without cut-off) was included
 in the model.}
\end{figure}
\subsection{Evolution of the scale height as a function of the distance to the center.}
To estimate the scale height corresponding to the different vertical profiles an exponential
 has been fitted to each profile. The result for the 3 fields is presented in Fig. 11 as a function
 of the distance to the Galactic center. It is clear that the 3 fields show a good consistency in
 the thickening of the profile. The effect seems to be somewhat larger at l$=$240, and is the 
 smaller at l$=$66. It is very hard to know what happens when going closer to the Galactic
 center, since our data do not go closer than 7.5 Kpc from the center. Unfortunately
 the lines of sight which would allow to probe the disk closer to the center are not
 in the areas released by the 2MASS team to date. 
\begin{figure} 
\centerline{\psfig{angle=0,figure=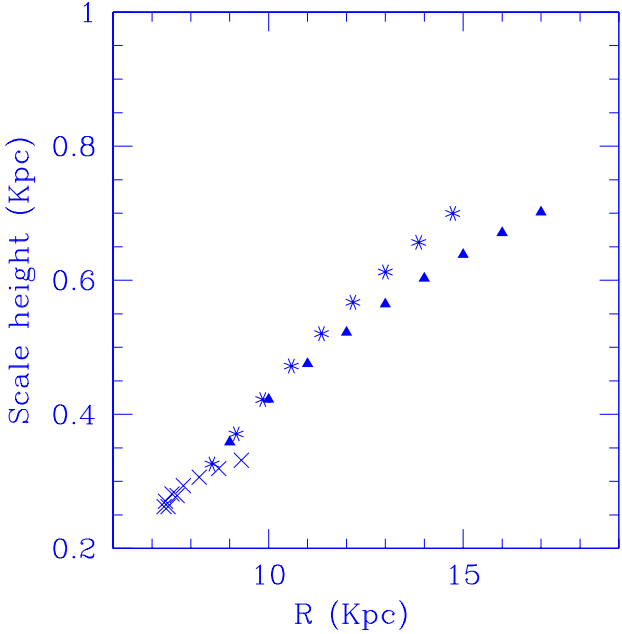,width=9cm}}
\caption{
 The variations of the scale height of the Z profile as a function of the distance
 to the Galactic center. The crosses represent the  field at l$=$66, the stars
 are for the field at l$=$240, and the triangle for the field at l$=$180.}
\end{figure}

%
%
%
%
\subsection{Full maximum likelihood fit.}
The former analysis conducted along each line of sight has been particularly 
useful to present the data, and demonstrate the existence of a basic feature: 
the thickening and flattening of the Z profile with distance. This effect has 
been well established, however it is also important to derive simple analytical
 full maximum likelihood solution of the whole data set. For this full maximum
 solution all the line of sight in the range $-50<b<50$ will be fitted at
 the same time. Note that has we did previously, we will fit by maximum likelihood, the decomposition
 in bins of magnitude (20 bins) of the distribution along the different line of sights,
 and not only the total counts.The effect of  disk thickening will be investigated in 2 the fields
 towards the outer disk, at l$=$180 and l$=$240. First, it is interesting to illustrate
 the fitting of different type of models to the data. To illustrate this issue we will
 present the result of the fitting of 4 different models for the field at l$=$180. 
 For all the models the luminosity function of Wainscoat {\it el al.} (1992) has been 
 used. The scale height for each population has also been taken from Wainscoat {\it el al.} 1992.\\\\
 \indent  {\bf 1-) Simple double exponential disk (Fig. 12):}
$$
 \rho(R,Z,l) = \exp \left ( -\frac{R-R_0}{R_H} -\frac{|Z-Z_0(R)|}{Z_H} \right)
$$
$$
{Z_0(R)} = \cases{  
 (R-R_0) \times A_W & R $>$ $R_0$ \cr
 & \cr
 0 & R $<$ $R_0$ \cr}
$$
 \indent  {\bf 2-) Double exponential disk + thick disk (Fig. 13):}
$$
 \rho(R,Z,l) = \exp \left ( -\frac{R-R_0}{R_H} -\frac{|Z-Z_0(R)|}{Z_H} \right) \ (1-R_{Sol}) 
$$
$$
+ \break  \exp{ \left ( -\frac{R-R_0}{R_{HT}} -\frac{|Z|}{Z_{HT}} \right)} \ R_{Sol}
$$
 \indent  {\bf 3-) Isothermal disk:}
$$
 \rho(R,Z,l) = \exp \left ( -\frac{R-R_0}{R_H}\right) \ {\rm sech^2} \left( -\frac{|Z-Z_0(R)|}{Z_H} \right)
$$
 \indent  {\bf 4-) Isothermal disk  + thick disk (Fig. 14):}
$$
 \rho(R,Z,l) = \exp \left ( -\frac{R-R_0}{R_H}\right) \ {\rm sech^2} \left( -\frac{|Z-Z_0(R)|}{Z_H} \right) \ (1-R_{Sol}) 
$$
$$
+ \break  \exp{ \left ( -\frac{R-R_0}{R_{HT}} -\frac{|Z|}{Z_{HT}} \right)} \ R_{Sol}
$$
 \indent  {\bf 5-) Disk with variable scale height (Fig. 15, l=180, Fig. 18, l=240):}
$$
 \rho(R,Z,l) = \exp \left ( -\frac{R-R_0}{R_H} -\frac{|Z-Z_0(R)|}{Z_H(R)} \right) \times (1+A_T)^{-1}
$$
$$
{Z_H(R)} = \cases{  
 \left(1+(R-R_0) \ A_T\right) \times H_Z  & R $>$ $R_0$ \cr
 & \cr
 1 & R $<$ $R_0$ \cr}
$$
 \indent  {\bf 6-) Disk with variable scale height and flattening (Fig. 17 l=180, Fig. 19, l=240):}
Let's define:
$$
 F(x,\beta) = 1-\exp \left( -\beta \ \exp(-x) \right)
$$
Then:
$$
 \rho(R,Z,l) = \exp \left ( -\frac{R-R_0}{R_H} \right) \times F \left ( \frac{|Z-Z_0(R)|}{Z_{HT}},\beta(R) \right) 
$$
$$
 \times \left( (1+A_T) \ \int_0^{+\alpha} F(x,\beta) dx \right)^{-1}
$$
With the same definition of $Z_H(R)$ as for the previous model, and:
$$
{\beta(R)} = \cases{  
 1+(R-R_0) \ A_F  & R $>$ $R_0$ \cr
 & \cr
 1 & R $<$ $R_0$ \cr}
$$
The best maximum likelihood fit for these 6 models are presented in Fig. 12,13,14,15,16,17.
One can note immediately that disk models with constant scale height do not give
 any acceptable fit of the data (Fig. 12 and 13). Even by including a thick disk 
 component, it is not possible to reproduce the star counts (Fig. 13). By 
 exploring different scale height scale length and ratio in the solar neighborhood,
 for the thick disk 
 one cannot reproduce the star counts properly. The situation is not improved, if 
 instead of an exponential profile, one adopt an isothermal profile for the Z  
 distribution (see Fig. 14, 15)
 However, as one can see in Fig. 16,
 just by adding a linear variation of the scale height with distance from the Sun,
 to the single exponential disk model, one can reproduce the star counts with much
 better accuracy. Note that this disk model with variable scale height show 
 this good agreement with the data, with even less free parameters than the 
 model with a thick disk component, which despite more parameters is not able
 to reproduce the star counts. An even better agreement is possible if one
 introduces some flattening of the disk profile by using a ``saturation'' parameter
 $\beta$ (see model (6) for the definition of $\beta$, and Fig. 16 for the fit).
 Finally, model (5) and (6) are fitted to the data in the field at l$=$240. 
 As one can check in Fig. 18 and 19, these simple model reproduce the data
 with quite good accuracy. Note that the thickening of the disk calculated using
 model (5) is the same for the 2 fields: $Z_H=(1+0.32 \ (R-R_0))\times H_Z$. It is
 also interesting to see that the flattening of the profile is larger at l$=$240,
 something that is also visible in our previous analysis. However something
 more troublesome is the fact that the scale length of the disk differs in
 the 2 fields. Some bias can be introduced by the presence of a spiral arm towards
 the anti-center, however
 the contrast in density of a spiral arm in the old population is small and unlikely
 to produce such discrepancies. This difference in scale heights is more likely due 
 to the fact that even if our model reproduces well the star counts, it is too
 simple to account for the complexity of the situation. The thickening of the disk
 is a flaring process and is likely to be non-linear, and the shape of the Z profiles
 are probably more complex than our simple analytical forms.
 To conclude with the full
 maximum likelihood fit, one can say that the results are in very good agreement with
 the simple analysis along each line of sight that was conducted before. The basic
 features, the thickening and flattening of the outer disk region is confirmed. 
\begin{figure} 
\centerline{\psfig{angle=0,figure=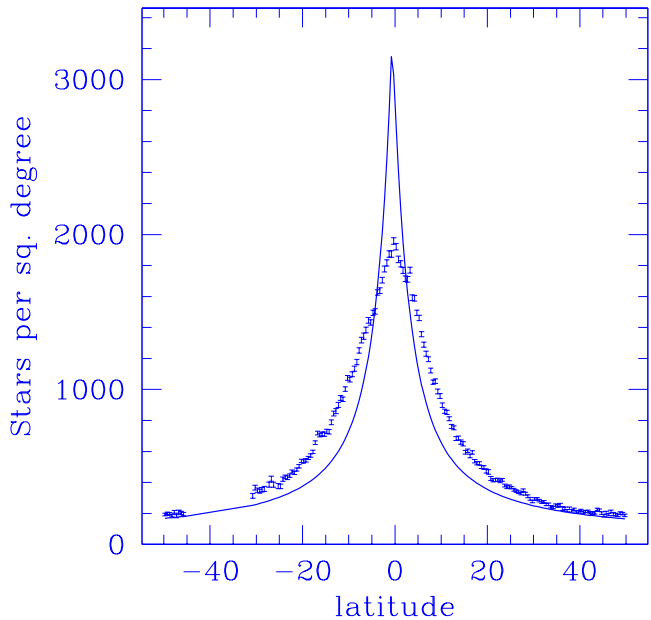,width=9cm}}
\caption{Fitting the data near l=180 using a single disk (double exponential)
 component. The parameters of the best fit are: $R_H=2.45$, $A_W=-0.007$.}
\end{figure}
\begin{figure} 
\centerline{\psfig{angle=0,figure=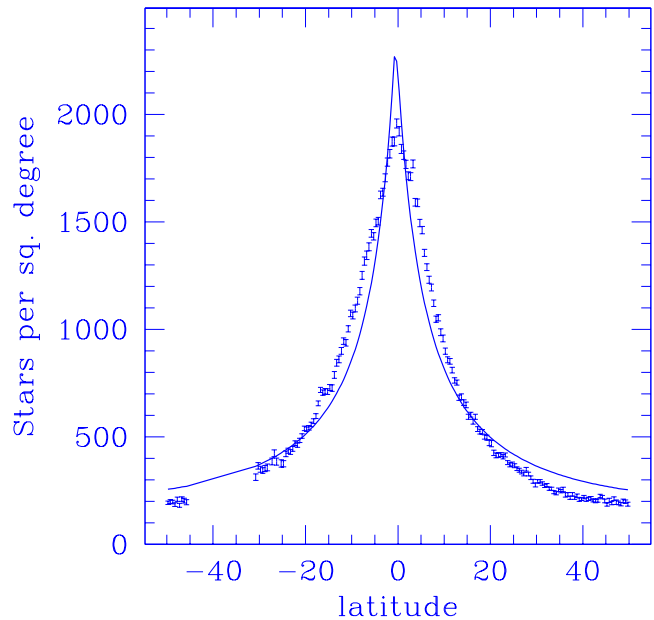,width=9cm}}
\caption{Fitting the data near l$=$180 using a disk + a thick disk 
 component. The parameters of the best fit are: $R_H=1.8$, $A_W=-0.007$, $R_{HT} =
 3.0$, $Z_{HT}=1.0$, $R_{Sol}=0.15$}
\end{figure}
\begin{figure} 
\centerline{\psfig{angle=0,figure=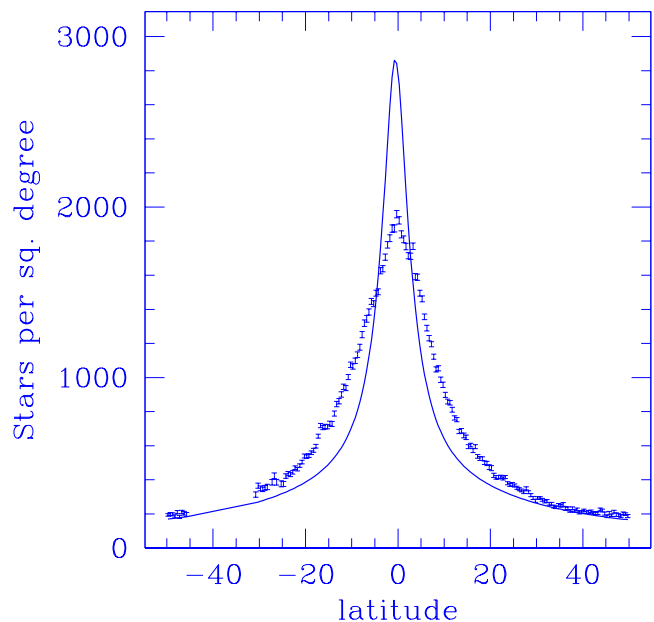,width=9cm}}
\caption{Single isothermal component (l$=$180). The parameters of the best fit are: $R_H=2.25$,
 $A_W=-0.007$}
\end{figure}
\begin{figure} 
\centerline{\psfig{angle=0,figure=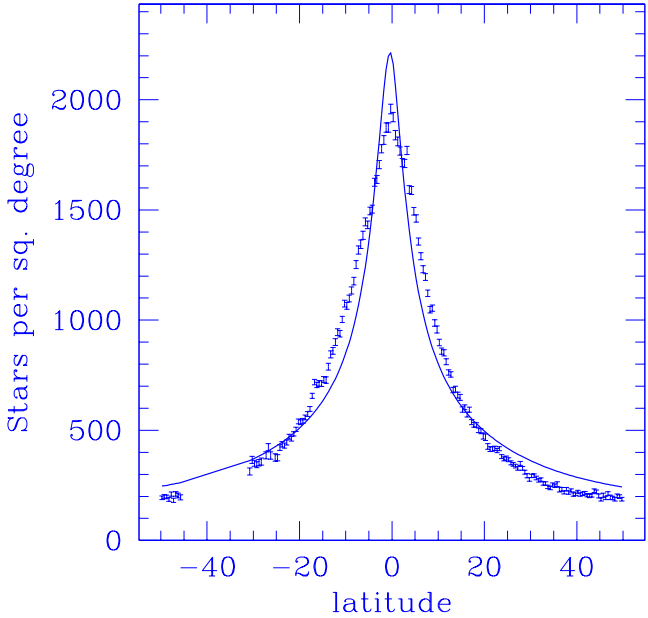,width=9cm}}
\caption{Fitting the data near l=180 using an isothermal + a thick disk 
 component. The parameters of the best fit are: $R_H=1.75$, $A_W=-0.007$, 
$R_{HT} = 3.2$, $Z_{HT}=1.0$, $R_{Sol}=0.13$}
\end{figure}
\begin{figure} 
\centerline{\psfig{angle=0,figure=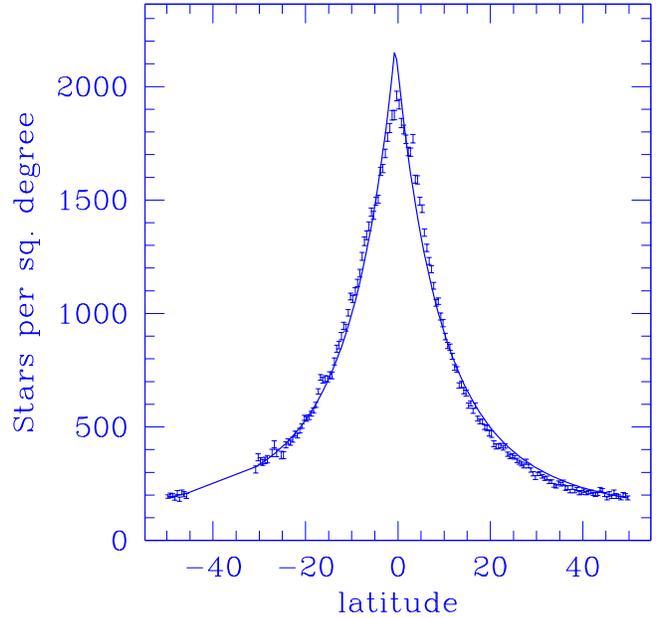,width=9cm}}
\caption{Fitting the data near l=180 using a disk with linearly variable scale height.
 The parameters of the best fit are: $R_H=3.05$, $A_W=-0.007$, $A_T=0.32$.}
\end{figure}
\begin{figure} 
\centerline{\psfig{angle=0,figure=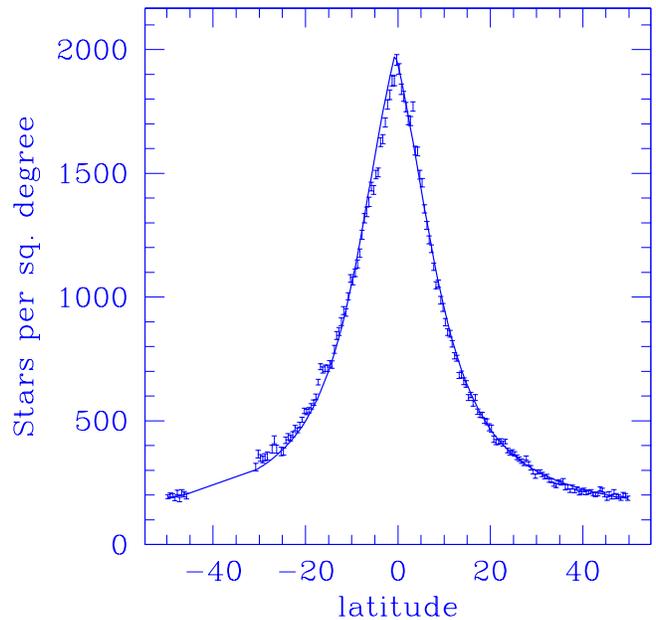,width=9cm}}
\caption{Fitting the data near l=180 using a disk with linearly variable scale height 
and flattening of the profile with distance. The parameters of the best fit are: $R_H=3.05$, 
$A_W=-0.007$, $A_T=0.15$, $A_F=0.6$.}
\end{figure}
\begin{figure} 
\centerline{\psfig{angle=0,figure=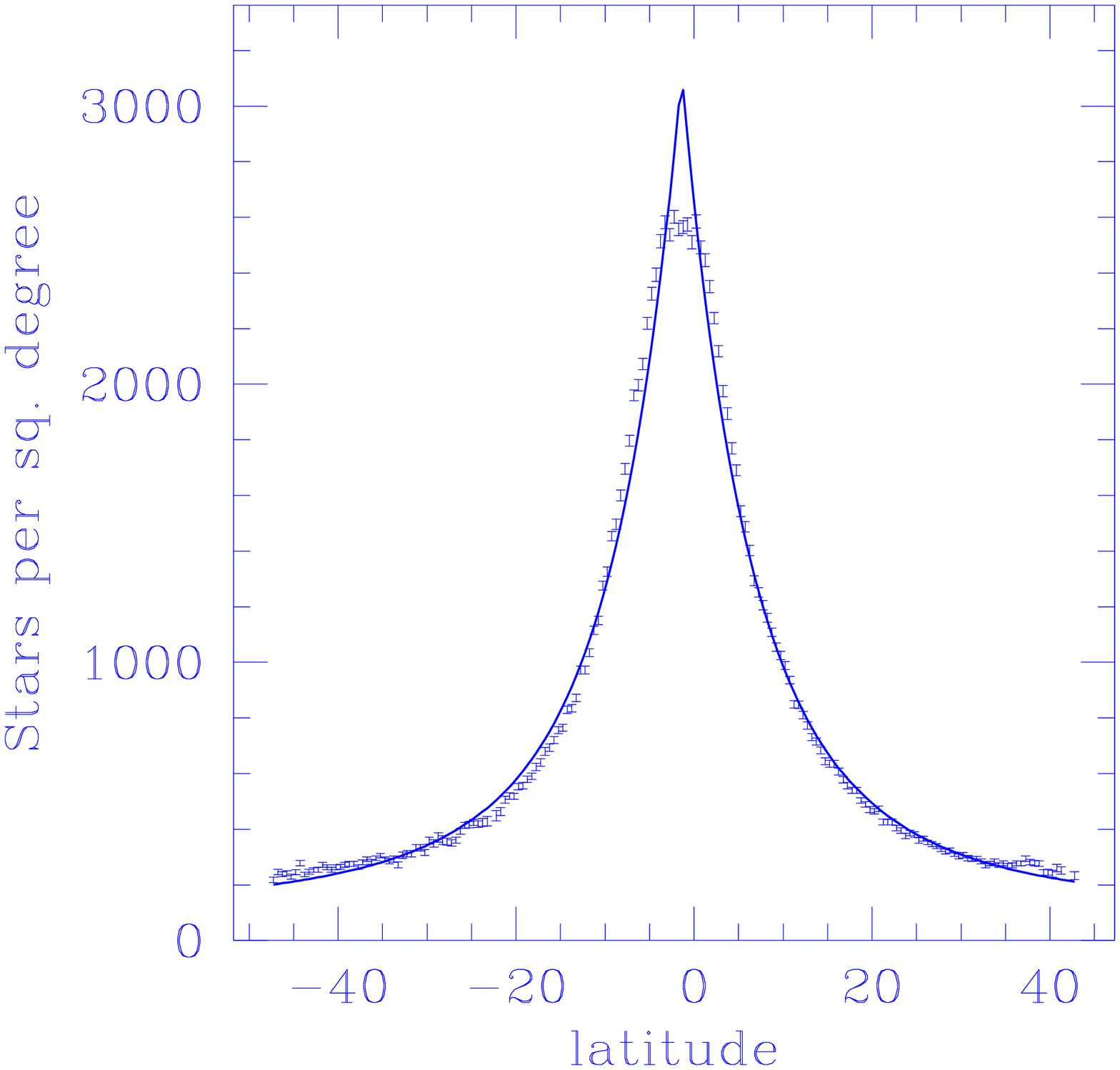,width=9cm}}
\caption{Fitting the data near l=240 using a disk with linearly variable scale height. 
The parameters of the best fit are: $R_H=2.45$, $A_W=-0.041$, $A_T=0.32$, $Z_{HT}=1.0$.}
\end{figure}
\begin{figure} 
\centerline{\psfig{angle=0,figure=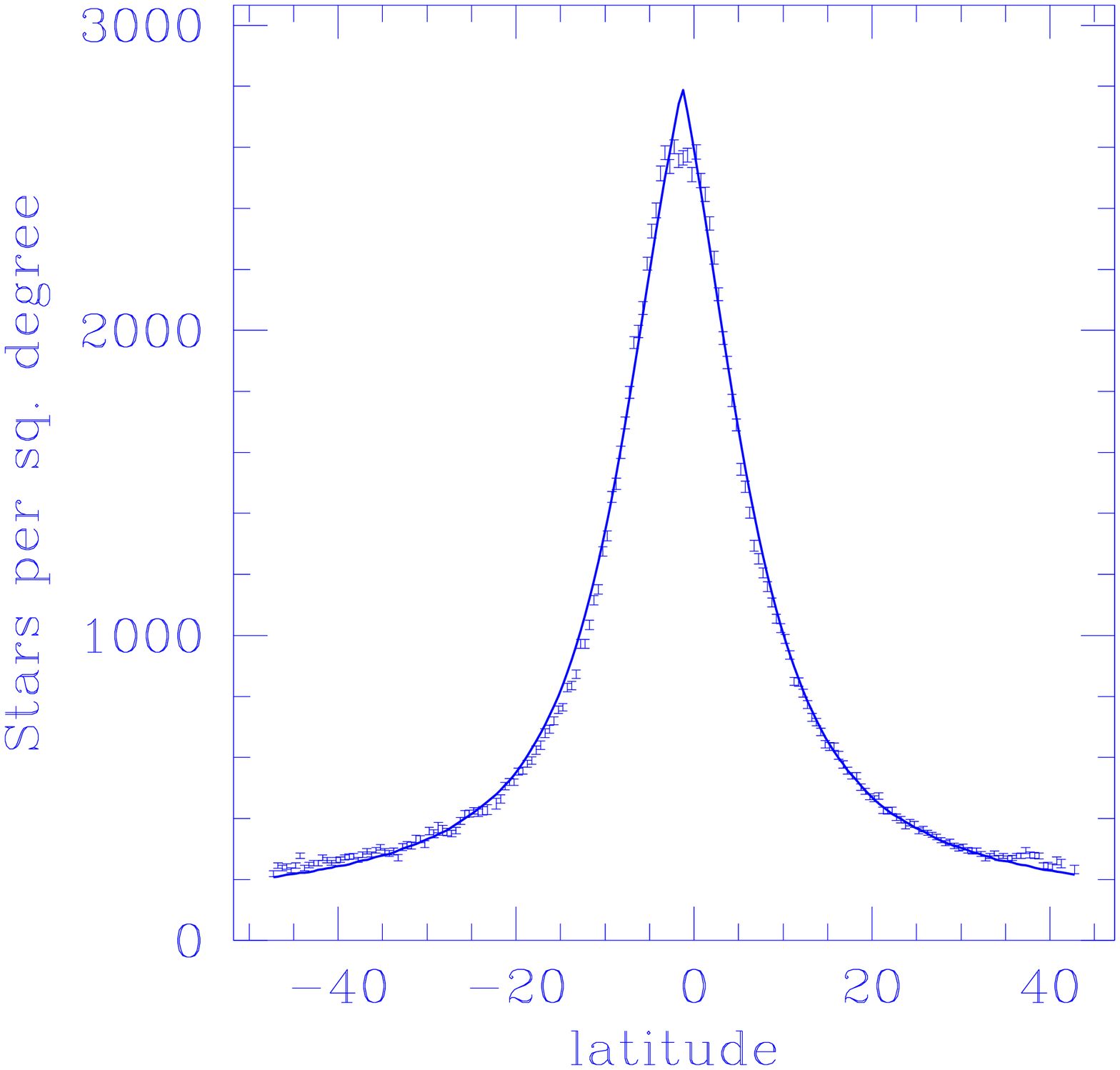,width=9cm}}
\caption{Fitting the data near l=240 using a disk with linearly variable scale height. 
The parameters of the best fit are: $R_H=2.45$, $A_W=-0.041$, $A_T=0.14$, $A_F=0.81$, $Z_{HT}=1.0$}
\end{figure}
\section{Discussion.}
 Our first analysis, which used a first order re-construction of the Z
 profile along each the line of sight has been very useful to present the data
 and describe the basic effect. Once the thickening of the disk  with distance had been
 established, a full maximum likelihood of the data has been performed. The full
 maximum likelihood analysis confirms the result obtained previously, and show
 that simple model, with linear variations of the scale height and flattening
 are in good agreement with the star counts. The maximum likelihood analysis 
 confirms that the warping of the disk is present in the field at l$=$240. 
 As for the thickness of the disk, a linear variation starting form the position
 of the Sun has been used to describe the warp. The maximum likelihood analysis
 shows also that a significant amount of warping is present in the field
 near l$=$180, however the amplitude of the effect is much smaller than in the field
 at l$=$240 (about 6 times smaller). In this analysis, a large number
 of models has been tested, models with thick disk, isothermal disk models, disks
 models with cut-off..., none of them is in agreement with the data set, except
 the models including thickening of the disk with increasing distance from the Sun.
 We arrive at the obvious that the thickening of the disk is the only possible
 explanation. Note that we did not took into account the metallicity effects,
 but since metallicity is expected to decrease towards the outer disk, and that
 stars becomes brighter with decreasing metallicity, the effect goes into
 the other direction. Since we observe a drop of the star counts in line of sights
 near the Galactic plane, it cannot be caused by a decreasing metallicity.
 This possibility of a flared and warped stellar disk has already
 been evaluated by Ewans {\it el al.} (1998) as a possible source of microlenses
 for sources in the LMC. This analysis demonstrates unambiguously for the first time 
 the flaring and the warping of the stellar disk, and gives an accurate description 
 of the effect. It is interesting to note that for other galaxies (edge-on spiral)
 some evidence of thickening of the disk
 vertical profile in the outer regions is also observed by  de Grijs {\it et al.} (1996)
 (see Fig. 2 in de Grijs 1996). 
 As a whole we see that a self-consistent picture of the stellar disk 
 emerge from this work: the stellar component is warped and flared, just like
 the HI component. The direction of the Z offset due to the warp is the same
 as in the HI maps presented by  Burton \& te Linkel Hekkert (1986). 
 The amplitude of the effect is consistent with the HI to an accuracy of 
 about 30 \%. One may not forget that our model of the warp is very simple,
 in the range of distance investigated. Note that warp effect was treated in the linear
 approximation; it is possible that as is observed
 for the gas, the warp effect is not linear, however this linear approximation
 is sufficient to reproduce the data. Concerning the thickening of the disk
 it is interesting to make a comparison to the flaring of the HI disk. 
 This comparison is presented in Fig. 20, scale height of the HI is taken
 from Wouterloot {\it et al.} (1990). It is clear that the amplitude
 of the thickening of the disk follow is comparable to what is observed in HI.
 This thickening of the disk, offers a natural explanation to the
 so called set of different ``disk cut-off'' observed for the Milky way 
 (Robin {\it et al.} 1992, Ruphy {\it et al.} 1996, Freudeinrech 1998). 
 due to the thickening, stars close to the plane are raised above the
 plane. This thickening gives the impression of a drop in the star counts
 at low latitudes, and fitting a cut-off in such density will
 give inconsistent results for different latitudes (as illustrated in Fig. 10).
 This effect  may also confuse the separation between the thick disk
 and disk component. The separation between this 2 components may prove
 very difficult. At least this work shows that clear evidences of the thick
 disk should be searched rather towards the inner regions of the Galaxy.
 Obviously the modelisation of the thickening and of the warping
 is somewhat degenerated, and
 it is clear that non linear thickening laws would fit the data also. However,
 in such case it is better to keep the simplest model that can fit the data.
 More accurate determination of the outer disk would require to overcome
 the degeneracy due to the unknown distances of the sources. Such result
 could be easily achieved by using contact binaries. The contact binaries
 are good distance indicators, and are found in large numbers (about 1\% of
 the main sequence stars are contact binaries). These stars are ideal
 tools to probe the structure of the outer disk, and some experiment
 should be conducted in the near feature.
\begin{figure} 
\centerline{\psfig{angle=0,figure=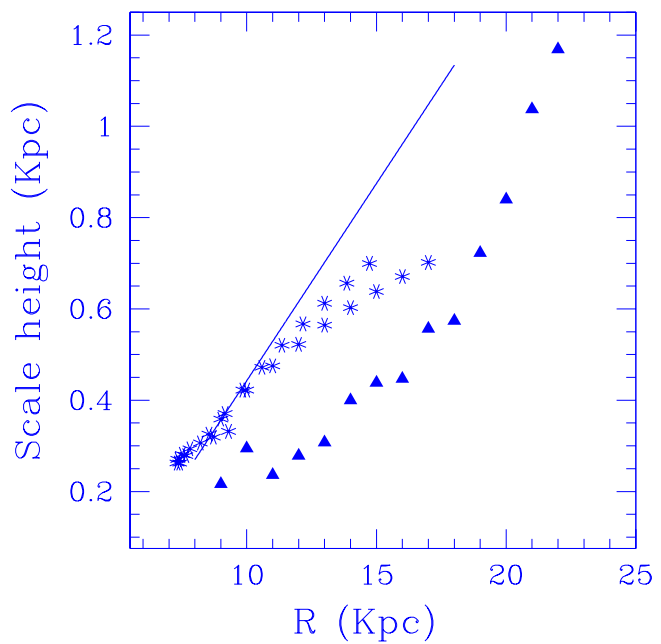,width=9cm}}
\caption{The distribution of the scale height of the stellar disk, -
 from first order reconstruction along the line of sight (stars) and from
 full maximum likelihood using model (5) (straight line). The full maximum
 likelihood solution as larger amplitude, this is due to the fact that
 model (5) does not include model flattening.}
\end{figure}
\label{lastpage}

\end{document}